\documentclass{iaus}
\usepackage{graphicx}

\title[Dead Zones and the Diversity of Exoplanetary Systems] 
{Dead Zones and the Diversity of Exoplanetary Systems}

\author[Yasuhiro Hasegawa \& Ralph E. Pudritz]   
{Yasuhiro Hasegawa$^1$
 \and Ralph E. Pudritz$^{1,2}$}

\affiliation{$^1$Department of Physics and Astronomy McMaster University, \\ Hamilton ON, 
L8S 4M1, Canada \\ email: {\tt hasegay@physics.mcmaster.ca} \\[\affilskip]
$^2$Origins Institute, McMaster University, \\ Hamilton ON,
L8S 4M1, Canada \\email: {\tt pudritz@physics.mcmaster.ca}}

\pubyear{2010}
\volume{276}  
\pagerange{001--002}
\jname{The Astrophysics of Planetary Systems: Formation, Structure, and Dynamical Evolution}
\editors{A. Sozzetti, M.G. Lattanzi \& A.P. Boss, eds.}
\begin{document}

\maketitle

\begin{abstract}
Planetary migration provides a theoretical basis for the observed diversity of exoplanetary systems.
We demonstrate that dust settling - an inescapable feature of disk evolution - gives even more rapid type I migration 
by up to a factor of about 2 than occurs in disks with fully mixed dust. On the other hand, type II migration becomes 
slower by a factor of 2 due to dust settling. This even more problematic type I migration can be resolved by the 
presence of a dead zone; the inner, high density region of a disk which features a low level of turbulence. 
We show that enhanced dust settling in the dead zone leaves a dusty wall at its outer edge. Back-heating 
of the dead zone by this wall produces a positive radial gradient for the disk temperature, which acts as a 
barrier for type I migration.  
\end{abstract}

\firstsection 
\section{Introduction}

Planetary migration is essential in the theory of planet formation in order to understand the observed 
mass-period relation (\cite[Udry \& Santos 2007]{UdrySantos07}). There are actually two types of migration, 
distinguished by planetary mass (e.g., \cite[Ward 1997]{Ward97}). Low-mass planets ($\sim$ several $M_{\oplus}$) 
undergo the so-called type I migration wherein angular momentum is transferred between planets and the surrounding 
gaseous disks only at the Lindbald and corotation resonances. In standard disk models, the planets efficiently 
lose their angular momentum, and hence plunge towards the central star within the disk lifetime 
($\sim 1-3 \times 10^6$ years). On the other hand, massive planets ($\sim 1 M_{J}$), known as type II migrators, 
can open up a gap in the disk due to strong, non-linear resonant torques, and migrate on the viscous evolution 
timescale of the disk. 

The long standing problem of migration is to identify what physical process(es) and/or condition(s) makes type I 
migration much slower. Considerable effort has been focused on disk properties such as density and temperature 
gradients to which the tidal torque is sensitive (\cite[Hasegawa \& Pudritz 2010a]{HasegawaPudritz10a}). 

\section{Dust Settling \& Rapid Planetary Migration}

Dust settling is observationally confirmed in disks around a variety of young stars (e.g., 
\cite[Hasegawa \& Pudritz 2010b]{HasegawaPudritz10b}). \cite[Hasegawa \& Pudritz 2010c]{HasegawaPudritz10c} first 
included the effects of this inescapable aspect of disk evolution on migration, by self-consistently computing 
the thermal structure of disks. In the computations, the full wavelength dependent, radiative transfer equation 
is solved by means of a Monte Carlo method, including dust settling and the gravitational force of a planet. 

We show that dust settling results in even more rapid type I migration, by up to a factor of about 2. This arises 
due to the geometrically flatter shape of the disk which is a consequence of dust settling. On the other 
hand, dust settling both lowers the gap-opening mass and slows the type II migration rate by about a factor of 2. 
This can be also understood as a consequence of flatter disk structures. It is obvious that some sort of more 
robust slowing mechanism is required for even more rapid type I migration.   

\begin{figure}[b]
\begin{center}
 \includegraphics[height=2.9cm]{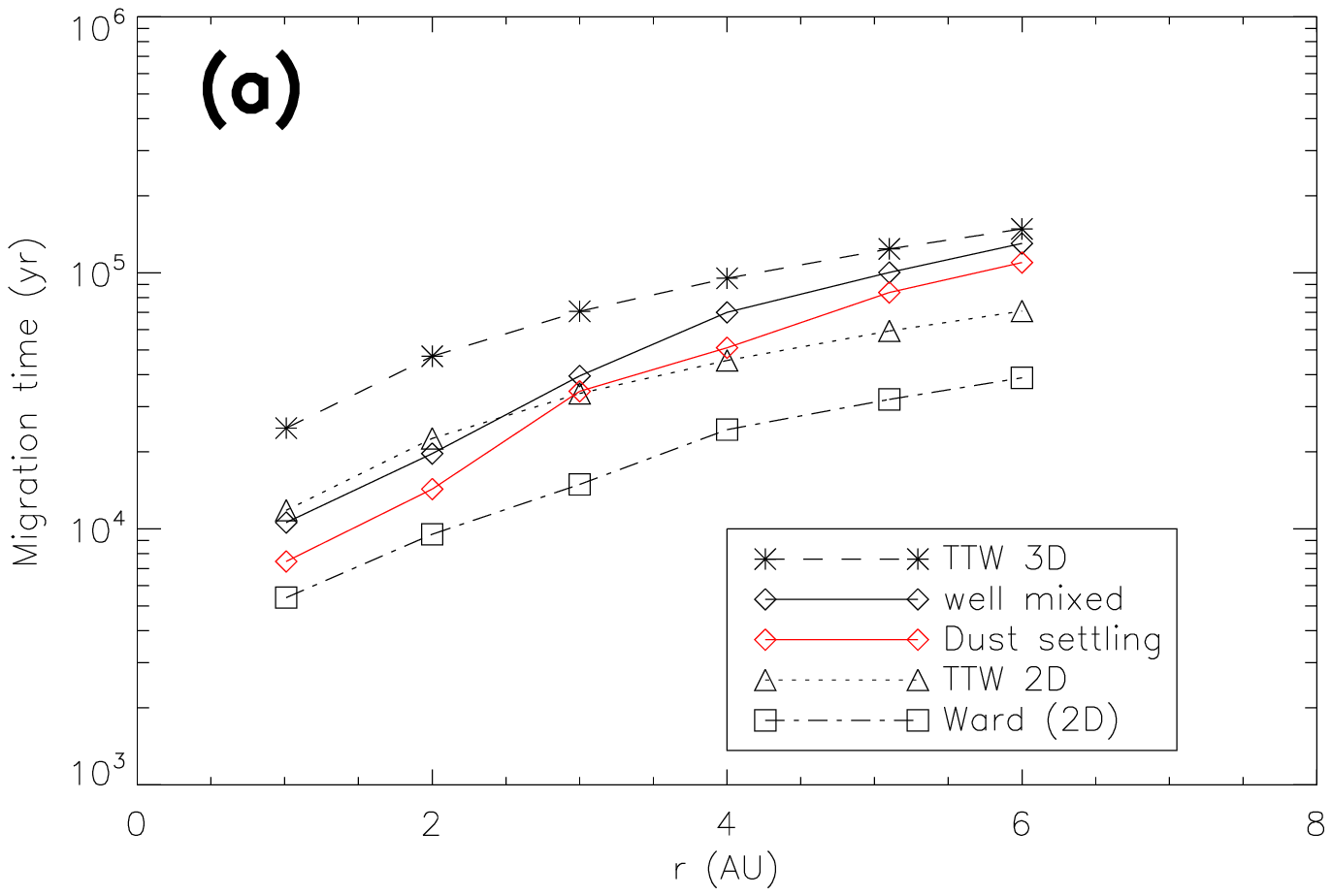}
 \includegraphics[height=2.9cm]{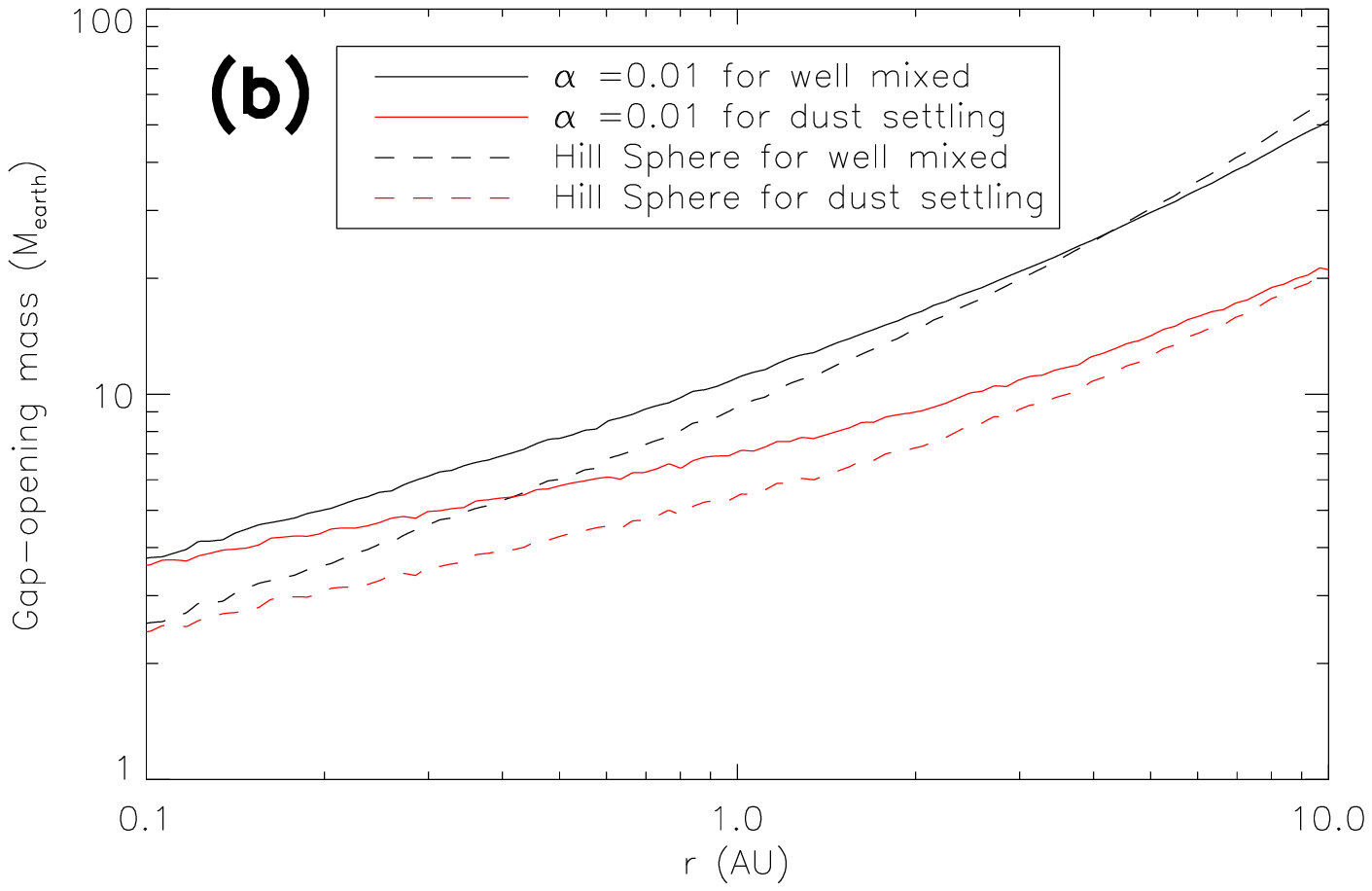}
 \includegraphics[height=2.9cm]{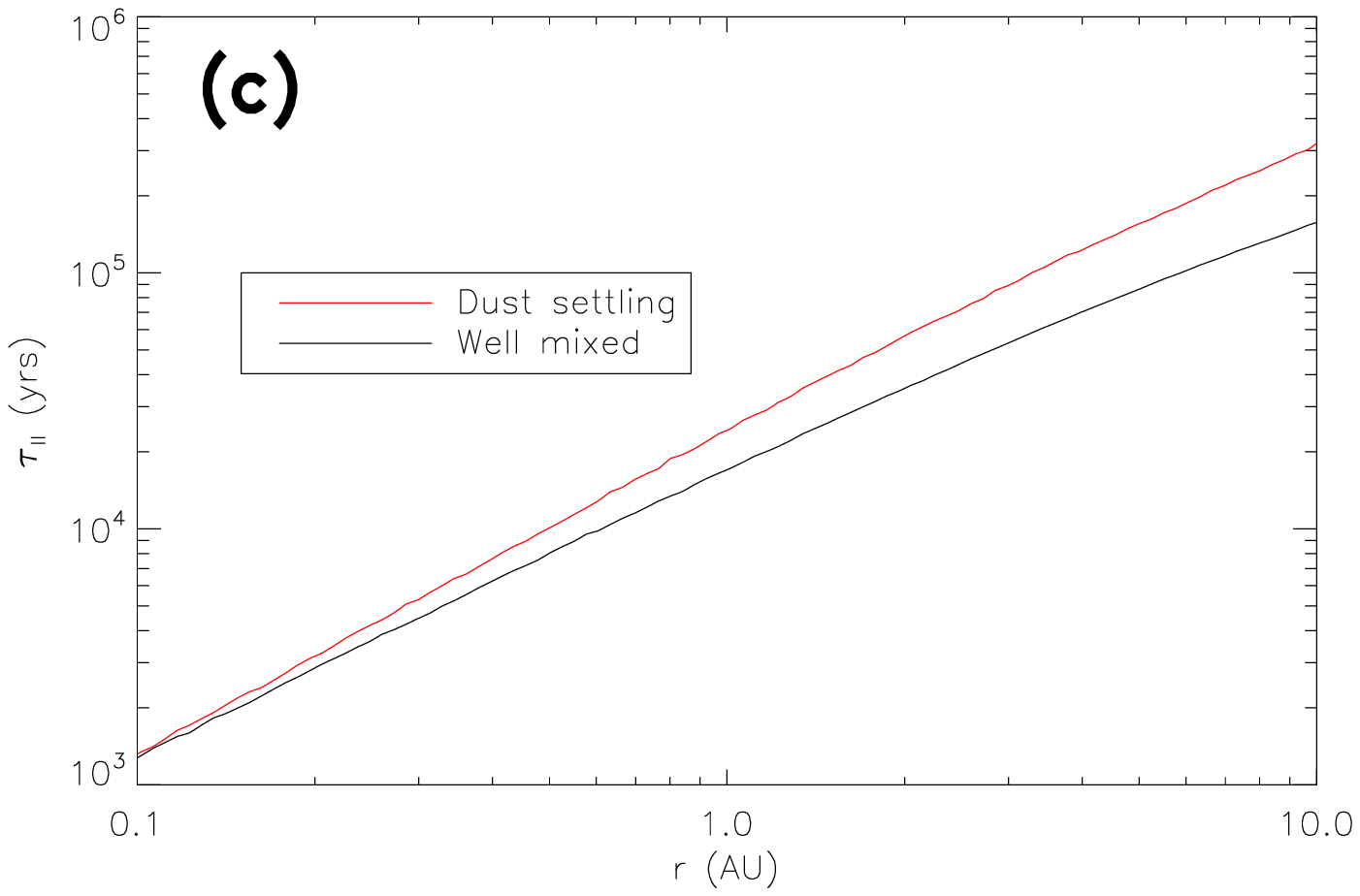}
 \includegraphics[height=2.1cm]{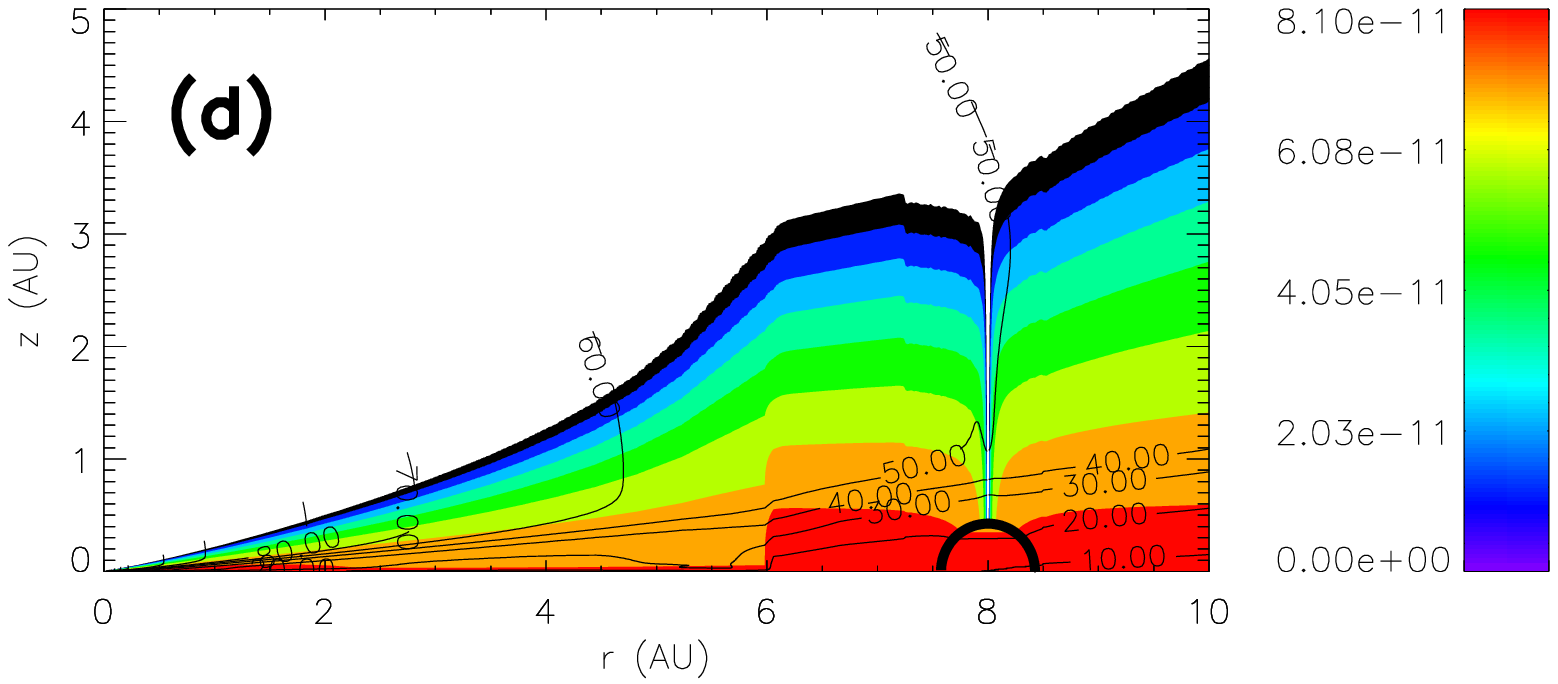}
 \includegraphics[height=2.1cm]{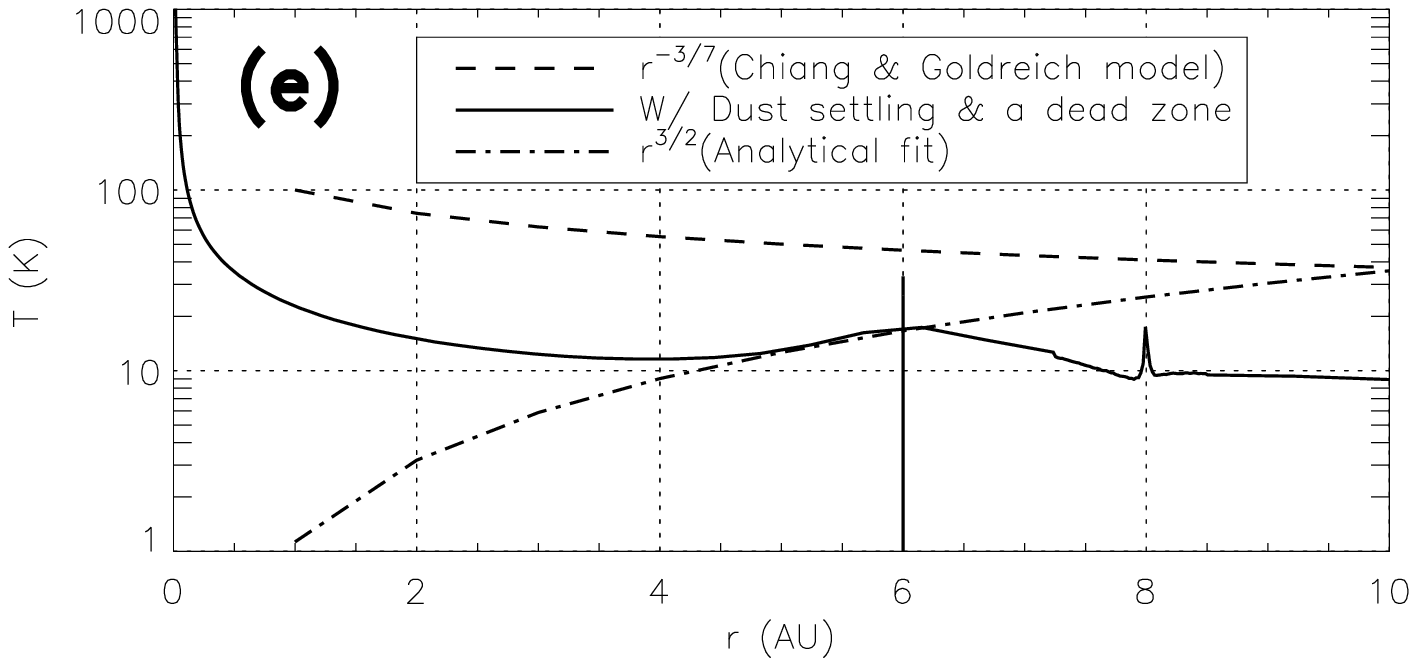}
 \includegraphics[height=2.9cm]{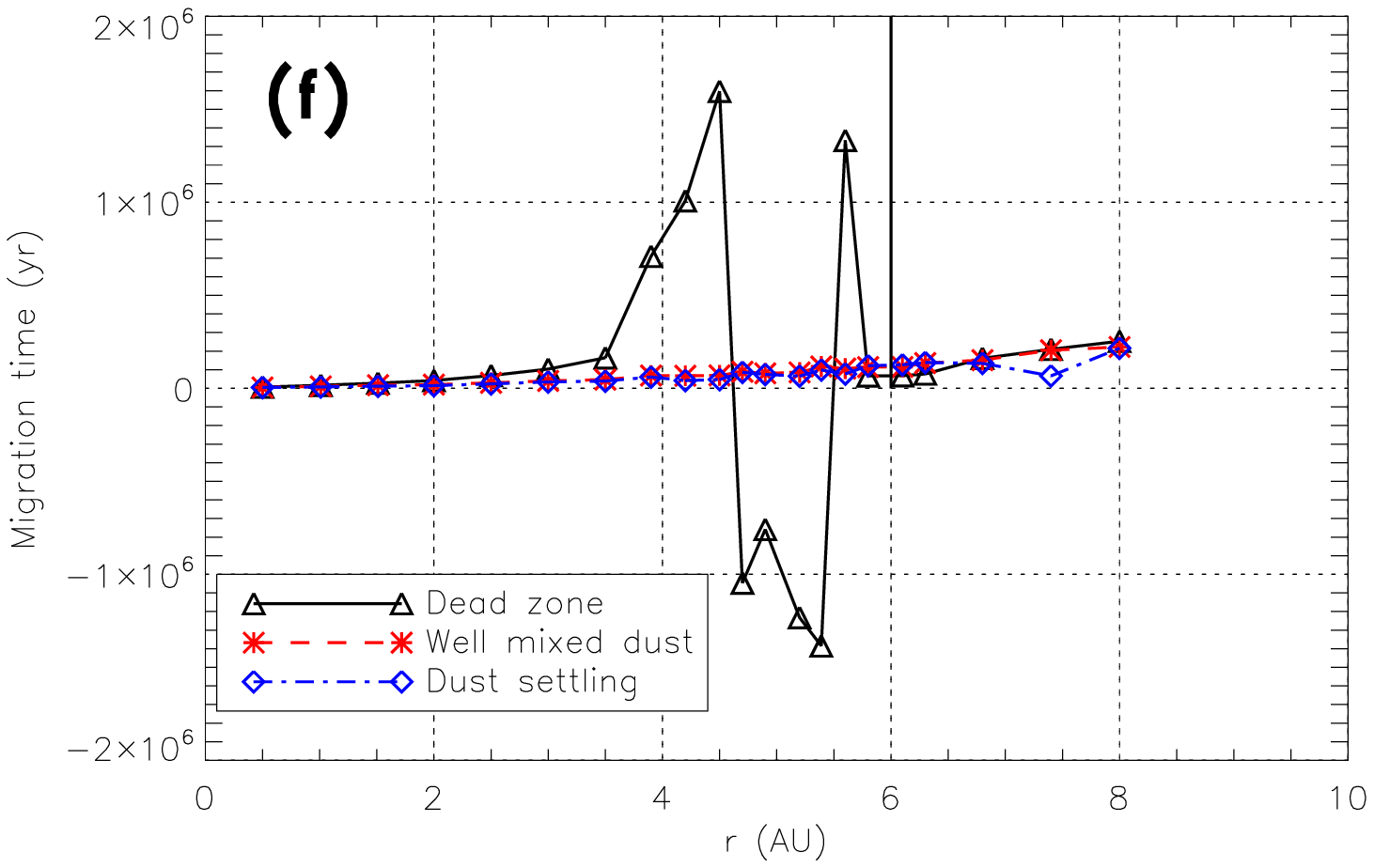} 
 \caption{Planetary migration in radiatively heated disks for the well mixed and dust settling cases (in (a)-(c); 
Adapted from \cite[Hasegawa \& Pudritz 2010c]{HasegawaPudritz10c}). In a), the migration time as a function of 
distance from the central star. In b), the gap-opening mass as a function of distance from the central star. In c), 
the timescale of type II migration. Planetary migration in radiatively heated disks with dead zones which is 6 AU in 
size (in (d)-(f); Adapted from \cite[Hasegawa \& Pudritz 2010a]{HasegawaPudritz10a}). In d), the dust density 
distribution with the temperature contours. In e), the temperature structure in the mid-plane region. In f), 
the migration time as a function of distance from the star.}
   \label{fig1}
\end{center}
\end{figure}

\section{Dead Zones \& Outward Migration}

Dead Zones change the structure of disks (\cite[Hasegawa \& Pudritz 2010a]{HasegawaPudritz10a}). 
They are the high density, low ionized regions, so that turbulence induced by the magnetorotational 
instability (MRI) is strongly suppressed (e.g., \cite[Gammie 1996]{Gammie96}). In a dead zone, dust settling is 
highly enhanced, so that a dusty wall is left at its outer boundary. The wall becomes thermally hot by stellar 
irradiation, and produces a positive temperature gradient due to its back-heating of the dead zone. We demonstrate 
that this temperature behavior results in outward migration (Fig. 1). We will address the role of this barrier in 
the mass-period relation by developing our population synthesis models.

\end{document}